# Digital Signature Security in Data Communication

[1]Robbi Rahim
[1]Department of Informatics, Institut Teknologi Medan, Indonesia
usurobbi85@zoho.com

[2]Andri Pranolo
[2]Department Informatics, Universitas Ahmad Dahlan, Indonesia
andri.pranolo@tif.uad.ac.id

[3]Ronal Hadi, [4]Rasyidah
[3,4]Department of Information Technology, Politeknik Negeri Padang, Indonesia
ronalhadi@polinpdg.ac.id, rasyidah@polinpdg.ac.id

[5]Heri Nurdiyanto
[5]Department of Informatics, STMIK Dharma Wacana, Metro Lampung, Indonesia
herinurdiyanto@gmail.com

[6]Darmawan Napitupulu
[6]Research Center for Quality System and Testing Technology, Indonesian Institute of Sciences, Indonesia
darwan.na70@gmail.com

[7]Ansari Saleh Ahmar
[7]Departement of Statistics, Universitas Negeri Makasar, Indonesia
ansarisaleh@unm.ac.id

[8]Leon Andretti Abdillah
[8]Departement of Information System, Universitas Bina Darma, Indonesia
leon.abdillah@yahoo.com

[9]Dahlan Abdullah
[9]Department Informatics, Universitas Malikussaleh, Indonesia
dahlan@unimal.ac.id

*Abstract*—Authenticity of access in very information are very important in the current era of Internet-based technology, there are many ways to secure information from irresponsible parties with various security attacks, some of technique can use for defend attack from irresponsible parties are using steganography, cryptography or also use digital signatures. Digital signatures could be one of solution where the authenticity of the message will be verified to prove that the received message is the original message without any change, Ong-Schnorr-Shamir is the algorithm are used in this research and the experiment are perform on the digital signature scheme and the hidden channel scheme.

## I. INTRODUCTION

Digital signatures are an authentication mechanism that enables the message maker to add code that acts as its signature and also allows the message recipient to test the authenticity and integrity of the message[1], [2]. Ong-Schnorr-Shamir scheme is one technique digitally process signatures of a message[3].

Ong-Schnorr-Shamir has two schemes, a digital signature scheme and a hidden channel scheme [3]. A digital signature scheme will form the digital signature of a message[2], [4], [5]. The verification process is performed on messages and digital signatures to check the authenticity and integrity of the message [6], when verification is successful, it means the message is still original and hasn't modified by other parties. The subliminal channel scheme is similar to the digital signature scheme[6]. The difference is the subliminal channel system has a decryption process that disguises the original message, this research attempts to explain in detail the workings of digital signature schemes and subliminal channels with the Ong-Schnorr-Shamir method.

The use of digital signature scheme and the hidden channel scheme as a process of checking messages when an exchange of information is anticipated of tapping information by third parties in communication.

## II. METHODOLOGY

### A. Crytography

Cryptography [7]–[10] is the study of secret writing with the aim that data communications can be encoded and decoded back to prevent other parties wanting to know the content, using certain codes and rules and using other methods so that only the rightful parties can know the actual content of the message, Fig. 1.

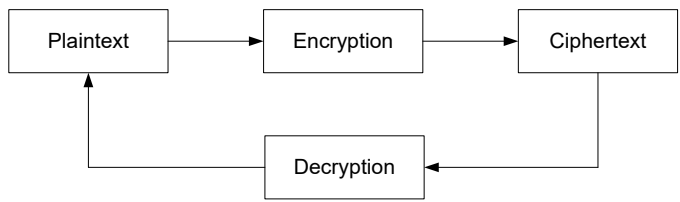

Fig. 1. Simple Cryptography Process

### B. Digital Signature

Digital signatures are an authentication mechanism that allows the message maker to add code that acts as its signature [2]. The signature is generated based on the message you want to sign and change according to the message. Digital signatures are sent together with a message to the recipient [2], [4].

Digital signatures enable the recipient of the information to test the authenticity of the information obtained and also to ensure that the data it receives is intact[3]. Therefore, public key digital signatures provide authentication and data integrity services [3]. Also, digital signatures also provide non-repudiation services, which means protecting the sender from a claim stating that he or she has sent information when there is no data transmission.

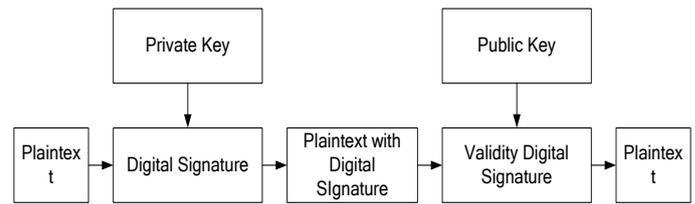

Fig. 2. Concept of Digital Signature

### C. The Ong-Schnorr-Shamir Scheme

The Ong-Schnorr-Shamir scheme [3] is one of the digital signature schemes contained in cryptography. The Ong-Schnorr-Shamir digital signature scheme was created by H.Ong, CPSchnorr, and A.Shamir and written in their book *An Efficient Signature Scheme Based on Polynomial Equations* on pages 208 to 216. This book was released to the public on the year 1984.

In addition to the digital signature scheme, Ong-Schnorr-Shamir also has a subliminal channel scheme (hidden channel). This scheme was created by Gustavus Simmons and was written in his book *The Prisoner's Problem and the Subliminal Channel* on pages 51 through pages 67 in 1984, "The Subliminal Channel and Digital Signatures" on pages 364 to page 378 in 1985 and " A Secure Subliminal Channel "on page 33 to page 41 in 1986 [11].

### D. Ong-Schnorr-Shamir Digital Signature Scheme

Working procedure of Ong-Schnorr-Shamir digital signature scheme [3].

1. Specify a large integer number (*n*) and an integer (*k*)
   a. *n* and *k* must be relatively prime, meaning the value of GCD (n, k) = 1
   b. N is the public key, meaning that other parties may know the value of n
   c. k is a private key, meaning that the sender only knows the value of k
2. Calculate the value of h with the following formula:
   $$h = -(k^{-1})2 \bmod n$$
3. Specify a random integer number (r)
   a. *n* and *r* must be relatively prime, meaning the value of GCD (n, r) = 1
   b. *r* is the public key, meaning the other *par*ty may know the value of r
4. Calculate S (1) and S (2) of the message (M) by using following formula:
   $$S_1 = 1/2 * \left(\frac{M}{r} + r\right) \bmod n$$
   $$S_2 = k/2 * \left(\frac{M}{r} - r\right) \bmod n$$
5. Verify messages and digital signatures using the following formula
   $$S_1^2 + h.S_2^2 \mid M (od\ n)$$

*E. Ong-Schnorr-Shamir Digital Signature Scheme*

Working procedure of Ong-Schnorr-Shamir Subliminal Channel scheme [3].

1. Specify a large integer number (*n*) and an integer (*k*).
    a. *n* and *k* must be relatively prime, meaning the value of GCD (n, k) = 1
    b. *n* is the public key, meaning that the value of *n* may be known by other parties
    c. *k* is a private key. The value of *k* is known by the message maker and the party that will decrypt the message
2. Calculate the value of h with the following formula.
$$h = -(k^{-1})2 \bmod n$$
3. Create original message (*w*), pseudo (*w′*) and count *S1* and *S2!*
    a. An incognito (*w′*) message was created to disguise the original message. The variable values *w*, *w′* and *n* must be relatively prime (*GCD* (*w′*, *n*) = 1 and *GCD* (*w*, *n*) = 1).
    b. S1 and S2 are signatures
    c. S1, S2 and w' are sent to the recipient
$$S_1 = 1/2 * (W'/w + w) \bmod n$$
$$S_1 = k/2 * (W'/w - w) \bmod n$$
4. Verification of pseudonyms and digital signatures (w ') using the following formula
$$W' = S_1^2 + h.S_2^2 (\bmod n)$$
5. Decrypted a pseudonym (w ') using the following formula:
$$W = \frac{W'}{S_1 + K^{-1} * S_2}$$

### III. PROPOSED METHOD

Ong-Schnorr-Shamir Digital Signature Scheme and Ong-Schnorr-Shamir Subliminal Channel Scheme experiment could be seen in the following process, the first test is to perform the process security message by using Ong-Schnorr-Shamir Digital Signature Scheme, see the experiment process below.

1. Bob choose $n = 239915931$ and $k = 658$.
2. Bob count of *h* value as:
   h = - (1/432964) mod 239915931
   h = - 0.000002309661
3. Bob choose r = 17.
4. Calculate $S_1$ and $S_2$ (*digital signature* from Bob)
   M = R = 82
   S (1) = 1/2 * (82/17 + 17) mod 239915931
   S (1) = 10.911764
   S (2) = 658/2 * (82/17 - 18) mod 239915931
   S (2) = -4006.0588
5. Alice verified message and signature from bob
   n = 239915931, h = -0.0000023, r = 17
   M = R = 82
   S (1) = 10.911764, S (2) = -4006.0588
   (10.911764)^2+ -0.0000023. (-4006.0588)^2 = 82
6. 82 = 82

The above process is to ensure Alice can verify Bob's digital signatures to ensure the authenticity and integrity of the message, Fig. 3 is the Ong-Schnorr-Shamir Digital Signature Scheme process diagram.

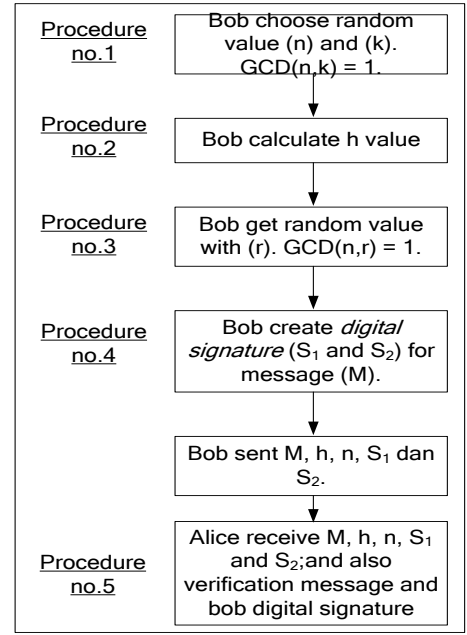

Fig. 3. Ong-Schnorr-Shamir Digital Signature Scheme process

The communication process described earlier is possible if there is no supervision made by a third party or communications may be made with the encrypted message, what if the communications conditions in observed area by a third party and also the message must not be in an encrypted status then the possible solution is to use Ong-Schnorr-Shamir Subliminal Channel Scheme can be seen in Fig. 4.

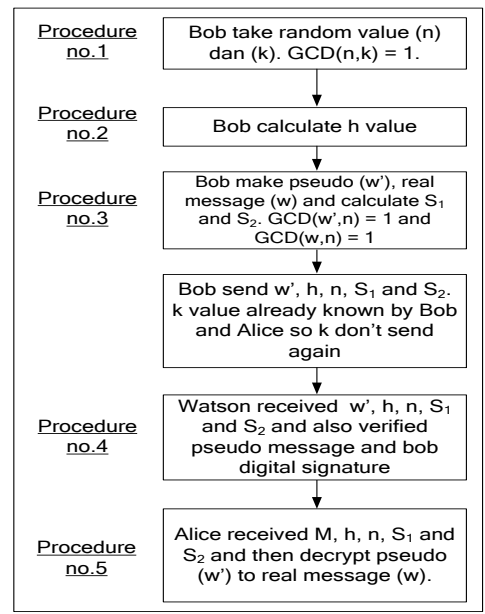

Fig. 4. Ong-Schnorr-Shamir Digital Signature Subliminal Channel Scheme

The message to be sent is 'R', see the process below for Ong-Schnorr-Shamir Subliminal Channel Scheme:

1. Bob choose $n$ = 17921593 dan $k$ = 421
2. Bob calculate $h$ value.
   $h$ = -(1/177241) mod 17921593
   $h$ = -0.000005642
3. Assume (w) = 'R' and pseudo messages (w') = 'A', then calculate S (1) and S (2) (*digital signature* from Bob)
   w = real message 'R' = 82
   w' = pseudo message 'A' = 65
   S (1) = 1/2 * (65/82 + 82) mod 17921593
   S (1) = 41.396341
   S (2) = 421/2 * (65/82 - 82) mod 17921593
   S (2) = -17094.140243
4. Watson verified signature and pseudo message (w') from Bob.
   w' = 'A' = 65
   S (1) = 41.396341, S(2) = -17094.140243
   w' = (41.396341)^2 + -0.000005642 . (-17094.140243)^2
   65 = 65 (True)
5. Alice decrypt pseudo message (w') to real message (w).
   w' = 'A' = 65
   S (1) = 41.396341, S(2) = -17094.140243
   w = 65 / (41. 396341 + -17094.140243/421)
   w = 85 (Char from ASCII 85 = 'R')

Communications bob and alice with conditions under surveillance and there should be no encrypted message, the Ong-Schnorr-Shamir Subliminal Channel algorithm can be used to communicate by creating false messages from the original message to transmitted and only known by senders and recipients, based on experiment conducted using Ong-Schnorr-Shamir Digital Signature Scheme and Ong-Schnorr-Shamir Subliminal Channel Scheme obtained the result that messages can be transmitted properly even under strict supervision conditions.

## IV. RESULTS AND DISCUSSION

The results of the Ong-Schnorr-Shamir Digital Signature Scheme and Ong-Schnorr-Shamir Subliminal Channel Scheme experiment are known to have different results, please see a few experiment in Table I.

TABLE I. RESULT ONG-SCHNORR-SHAMIR DIGITAL SIGNATURE SCHEME

| No | Message | Digital Signature |
|---|---|---|
| 1 | Robbi | <begin_of_signature><br>3093.00662786938\|-2901227.78305852<br>3093.00897187197\|-2901225.58438409<br>3093.00792111219\|-2901226.56999677<br>3093.00792111219\|-2901226.56999677<br>3093.00848690592\|-2901226.03928225<br><end_of_signature> |
| 2 | Siang | <begin_of_signature><br>3093.00670869706\|-2901227.70724216<br>3093.00848690592\|-2901226.03928225<br>3093.00784028451\|-2901226.64581313<br>3093.00889104429\|-2901225.66020045<br>3093.00832525057\|-2901226.19091497<br><end_of_signature> |
| 3 | Nomor HP berapa | <begin_of_signature><br>3093.00630455868\|-2901228.08632396<br>3093.00897187197\|-2901225.58438409<br>3093.00881021662\|-2901225.73601681<br>3093.00897187197\|-2901225.58438409<br>3093.009214355\|-2901225.35693501<br>3093.00258648561\|-2901231.5738765<br>3093.00581959263\|-2901228.54122212<br>3093.00646621403\|-2901227.93469124<br>3093.00258648561\|-2901231.5738765<br>3093.00533462658\|-2901228.99612027<br>3093.00816359522\|-2901226.34254769<br>3093.009214355\|-2901225.35693501<br>3093.00784028451\|-2901226.64581313<br>3093.00905269964\|-2901225.50856773<br>3093.00784028451\|-2901226.64581313<br><end_of_signature> |
| 4 | No Rekening BNI | <begin_of_signature><br>3093.00630455868\|-2901228.08632396<br>3093.00897187197\|-2901225.58438409<br>3093.00258648561\|-2901231.5738765<br>3093.00662786938\|-2901227.78305852<br>3093.00816359522\|-2901226.34254769<br>3093.00864856127\|-2901225.88764953<br>3093.00816359522\|-2901226.34254769<br>3093.00889104429\|-2901225.66020045<br>3093.00848690592\|-2901226.03928225<br>3093.00889104429\|-2901225.66020045<br>3093.00832525057\|-2901226.19091497<br>3093.00258648561\|-2901231.5738765<br>3093.00533462658\|-2901228.99612027<br>3093.00630455868\|-2901228.08632396<br>3093.0059004203\|-2901228.46540575<br><end_of_signature> |
| 5 | Password | <begin_of_signature><br>3093.00646621403\|-2901227.93469124<br>3093.00784028451\|-2901226.64581313<br>3093.00929518267\|-2901225.28111866<br>3093.00929518267\|-2901225.28111866<br>3093.00961849337\|-2901224.97785322<br>3093.00897187197\|-2901225.58438409<br>3093.009214355\|-2901225.35693501<br>3093.00808276754\|-2901226.41836405<br><end_of_signature> |

Experiment with Ong-Schnorr-Shamir Digital Signature Scheme based on the steps and functions already described, the verification process is complete only to test the message that was sent whether original and not modified, for simulation process see Fig. 5.

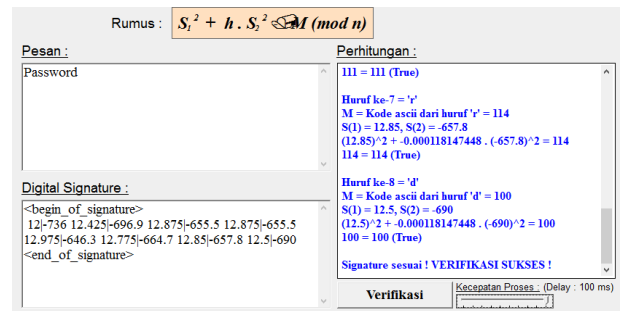

Fig. 5. Simulation Ong-Schnorr-Shamir Digital Signature Scheme

Ong-Schnorr-Shamir Subliminal Channel Scheme experiment results can be seen in Table II.

TABLE II. RESULT ONG-SCHNORR-SHAMIR SUBLIMINAL CHANNEL SCHEME

| No | Message | Message Disguise | Digital Signature |
|---|---|---|---|
| 1 | Robbi | Janner | <begin_of_signature> 41.451219512195\|-17800.9146341463 55.936936936937\|-24172.6846846847 49.561224489796\|-21264.6224489796 49.561224489796\|-21264.6224489796 52.980952380952\|-22836.3619047619 17.78125\|-6242.03125 <end_of_signature> |
| 2 | Siang | Janner | <begin_of_signature> 41.94578313253\|-18022.8012048193 52.961904761905\|-22844.7238095238 49.067010309278\|-21042.5824742268 55.5\|-23925.5 51.990291262136\|-22393.2621359223 17.78125\|-6242.03125 <end_of_signature> |
| 3 | Nomor HP berapa | Janner | <begin_of_signature> 39.474358974359\|-16912.7564102564 55.936936936937\|-24172.6846846847 55.004587155963\|-23703.9862385321 55.995495495496\|-24146.9774774775 57.44298245614\|-24828.5307017544 17.78125\|-6242.03125 36.222222222222\|-15706.4444444444 40.2\|-17472.2 16.5\|-6804.5 49.163265306122\|-21439.3265306122 50.658415841584\|-22099.9554455446 57.140350877193\|-24961.3859649123 48.664948453608\|-21219.087628866 56.142857142857\|-24521.2857142857 48.664948453608\|-21219.087628866 <end_of_signature> |
| 4 | No Rekening BNI | Janner | <begin_of_signature> 39.474358974359\|-16912.7564102564 55.936936936937\|-24172.6846846847 17.71875\|-6269.46875 41.670731707317\|-17704.5487804878 51\|-21950 54.032710280374\|-23252.6401869159 50.658415841584\|-22099.9554455446 55.145454545455\|-24081.1454545455 52.652380952381\|-22980.6047619048 55.145454545455\|-24081.1454545455 51.655339805825\|-22540.3058252427 16.5\|-6804.5 33.242424242424\|-14380.5757575758 39.205128205128\|-17030.9487179487 36.719178082192\|-15927.2808219178 <end_of_signature> |
| 5 | Password | Janner | <begin_of_signature> 40.4625\|-17356.9625 49\|-21072 57.978260869565\|-25032.5434782609 57.978260869565\|-25032.5434782609 59.924369747899\|-25934.2016806723 56.013513513514\|-24139.0675675676 57.140350877193\|-24961.3859649123 50.16\|-21879.76 <end_of_signature> |

Table II shows the results of communication done by giving a disguise message to cover the original message, the first process is done to verify the message whether the guise message is true or not and then proceed in the second stage to decryption process by execute disguise message with digital signature, see Fig. 6.

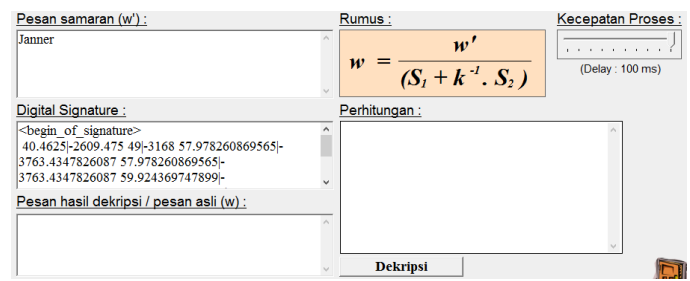

Fig. 6. Decryption process

The results of tests performed based on the security process, made that the Ong-Schnorr-Shamir Subliminal Channel Scheme algorithm is better and safer even though the messages sent (aliases) are known by third parties and will not arouse suspicion.

## V. CONCLUSION

Research experiment Ong-Schnorr-Shamir Digital Signature Scheme found that the Ong-Schnorr-Shamir Digital Signature Scheme can be used to maintain authentication and data integrity while on the Ong-Schnorr-Shamir Subliminal scheme Channel is a cryptographic method that can be used to disguise the original message and also this scheme supports the verification process of the Ong-Schnorr-Shamir Digital Signature scheme. Concerning security, it can conclude that Ong-Schnorr-Shamir Subliminal Channel is better than Ong-Schnorr-Shamir digital signature.